\documentclass[3p,sort&compress]{elsarticle}

\usepackage{lineno,amsmath,amsfonts,amssymb} 
\usepackage{tikz}
\usetikzlibrary{plotmarks}
\usetikzlibrary{arrows}
\usetikzlibrary[patterns]
\usetikzlibrary{shapes.geometric,arrows}

\usepackage{lineno}

\usepackage{soul}
\soulregister\cite7
\soulregister\ref7
\soulregister\eqref7

\usepackage{hyperref}
\hypersetup{
    colorlinks=true,
    linkcolor=blue,
    filecolor=magenta,      
    urlcolor=cyan,
}

\newcommand{\vect}[1]{\boldsymbol{#1}}

\journal{Acta Materialia}

\begin{document}

\begin{frontmatter}


\title{Enhanced mobility of dislocation network nodes and its effect on dislocation multiplication and strain hardening}

\author[llnl]{Nicolas Bertin\corref{cor1}}
\ead{bertin1@llnl.gov}

\author[stanford]{Wei Cai}
\author[llnl]{Sylvie Aubry}
\author[llnl]{Athanasios Arsenlis}
\author[llnl]{Vasily V. Bulatov}

\cortext[cor1]{Corresponding author}
\address[llnl]{Lawrence Livermore National Laboratory, Livermore, CA, USA}
\address[stanford]{Department of Mechanical Engineering, Stanford University, Stanford, CA, USA}

\begin{abstract}
Understanding plastic deformation of crystals in terms of the fundamental physics of dislocations has remained a grand challenge in materials science for decades. To overcome this, the Discrete Dislocation Dynamics (DDD) method has been developed, but its lack of atomistic resolution leaves open the possibility that certain key mechanisms may be overlooked. By comparing large-scale Molecular Dynamics (MD) with DDD simulations performed under identical conditions we uncover significant discrepancies in the predicted strength and microstructure evolution in BCC crytals under high-strain rate conditions. These are traced to unexpected behaviors of dislocation network nodes forming at dislocation intersections, that can move in ways not previously anticipated as revealed by MD. Once these newfound freedoms of nodal motion are incorporated, DDD simulations begin to closely match plastic evolution observed in MD. This additional mechanism of motion whereby non-screw dislocations can change their glide plane profoundly affects fundamental processes of dislocation multiplication, recovery and storage that define strength of metals.
\end{abstract}

\begin{keyword}
Dislocation multiplication \sep dislocation network nodes \sep dislocation mobility \sep dislocation cross-slip \sep Molecular Dynamics \sep Discrete Dislocation Dynamics
\end{keyword}

\end{frontmatter}


\section{Introduction} \label{sec:Intro}


Over nearly 90 years since their initial discovery, lattice dislocations have been firmly established as the main agents of crystal plasticity and yet quantitative prediction of macroscopic crystal plasticity directly from dislocation behavior remains a challenge \cite{van2020roadmap}. Computational method of Discrete Dislocation Dynamics (DDD) is widely viewed as a promising approach in which motion of each individual dislocation as well as interactions among dislocation lines are explicitly accounted for and thus directly define the overall crystal plasticity response \cite{kubin1992dislocation, zbib1998plastic, ghoniem2000parametric, weygand2002aspects, BulatovCai06, Arsenlis07}. Given the drastic reduction in the degrees of freedom -- from all atoms to just dislocation lines -- the DDD method has the promise to reach length and time scales relevant for crystal plasticity that are inaccessible to fully atomistic simulations. 
As a method bridging between microscopic dislocation theory and macroscopic crystal plasticity, DDD still faces serious computational limitations. However, algorithmic advances coupled with 
steadily growing computing capacity are bringing about DDD simulations on ever-increasing length- and time-scales \cite{bertin2020frontiers}.

Here we focus on a challenge altogether different from the method's computational efficiency, namely the limited or unknown physical fidelity of mechanisms of dislocation behavior presently included in DDD models. Being a mesoscopic approach, a DDD simulation only includes mechanisms that the developer knew about and chose to account for in the model formulation.
But what if one or several mechanisms of dislocation behavior are not known \emph{a priori} and thus never included in the DDD model? 

Historically, dislocation theory has been concerned with the behavior of individual dislocations, including mechanisms of dislocation mobility, dislocation core structure, cross-slip, climb, etc. Until first TEM observations of dislocations in the late 1950's, understanding of dislocation behaviors was based to a great extent on physical intuition and deductive reasoning. Yet, despite subsequent impressive developments in imaging techniques, including {\it in situ} transmission electron microscopy (TEM), experiments do not resolve dislocations in details sufficient to confirm some previously hypothesized mechanisms or to discover unknown atomistic mechanisms of dislocation motion. Since the 1960's, atomistic simulations of individual dislocations and, subsequently, small groups of dislocations have been increasingly used as means of inquiry into dislocation behaviors augmenting experiment. Presently, direct MD simulations performed at the limits of super-computing are reaching previously unachievable scales of simultaneous motion and interactions of thousands of dislocation lines statistically representative of macroscopic crystal plasticity at deformation rates of the order  $10^5$ s$^{-1}$ and higher \cite{Zepeda17, zepeda2021atomistic, bertin2023crystal}.

Equally important as their scales is that such direct MD simulations are fully atomistically resolved so that every feature in a simulated stress-strain curve can be unambiguously connected to the underlying dynamic events in the life of dislocations. In tandem with the recently developed accurate and efficient methods for dislocation extraction and indexing (DXA) \cite{Stukowski10, Stukowski14}, direct MD simulations now serve as a powerful {\it in silico} computational microscope. Unlike the more traditional atomistic simulations invariably probing behaviors of single dislocations or small groups of dislocations in configurations {\it presumed relevant} for crystal plasticity, in massive MD simulations one observes how dislocations collectively and naturally respond {\it en masse} to applied straining. Unbiased by human intuition, such simulations can reveal previously unanticipated mechanisms of dislocation behavior.

Direct MD simulations of crystal plasticity are especially informative when contrasted against DDD simulations performed under identical conditions, a practice we will refer to as cross-scale (X-scale) matching.
In this paper we present one example where X-scale matching bears fruit by exposing glaring discrepancies between MD and DDD predictions that are traced to a distinct type of dislocation network nodes and their modes of evolution that have not been previously considered. Colloquially referred to as \emph{sticky} hereafter, such nodes are immobile and restrict further motion of dislocation lines in DDD simulation. In contrast, in MD simulation the same sticky nodes can dissociate into mobile nodes thus preventing formation of dense dislocation tangles and excessive strain hardening. Discovery and kinematics of sticky nodes via topological rearrangement are the main focus of this paper.
We further show that, once the physics missing in DDD is added, its predictions fall close in line with corresponding MD simulations precisely where the two previously disagreed.

\section{Computational methods} \label{sec:methods}


We employ the X-scale matching approach whereby simulations of metal plasticity are performed using MD and DDD simulations by subjecting model single crystals of BCC tantalum (Ta) to the same loading conditions on the same length and time scales where both methods overlap.
The mesoscale approach of DDD has gained widespread recognition as a successful method for materials simulations, yet how well it reflects the underlying atomistic dynamics in strained crystals remains largely unknown \cite{bertin2020frontiers}. In the context of this study, we regard MD simulations as the ground truth for which we wish the DDD model to be a faithful proxy.
Here we demonstrate how direct one-to-one comparisons of dislocation trajectories initiated from the same configurations are used to assess the differences between the MD and DDD predictions and help us identify previously overlooked or missing physical mechanisms. Once uncovered, these mechanisms can be included as new rules in the DDD model to enable better agreement with MD predictions.

\subsection{MD simulations} \label{sec:MD}

MD simulations were performed in LAMMPS \cite{LAMMPS} using a previously reported interatomic potential for Ta \cite{Li03}.
Large-scale MD simulations were performed in a crystal volume containing $\sim 33$ million atoms which was determined in our previous work \cite{Zepeda17, STIMAC2022, bertin2022sweep} to be sufficient for statistically representative simulations of single crystal plasticity under compression at a rate of $2 \times 10^8$/s. Twelve hexagon-shaped prismatic dislocation loops of vacancy type were seeded at random locations in an initially perfect BCC crystal \cite{Zepeda17}.
After initial annealing, the crystals were subjected to uniaxial compression along the [001] crystallographic axis at a strain rate of $2 \times 10^8$/s. Uniaxial stress conditions and constant 300K temperature were maintained using the {\it langevin} thermostat and {\it nph} barostat. Dislocation networks were extracted along the MD trajectories using the DXA algorithm \cite{Stukowski10, Stukowski14} at time intervals ranging from 0.1 to 1 ps between snapshots.

Elemental dislocation networks containing sticky nodes with degree three (3-nodes) were created to verify the new mechanisms of 3-node motion in \S\ref{sec:network}. Full 3D periodic boundary conditions (3D-PBC) are convenient for our purposes as they enable relatively straightforward and accurate control of applied stress and eliminate unwanted boundary effects while preserving translational invariance. The price one pays for using 3D-PBC is that a minimum of two dislocation networks have to be inserted for the net Burgers vector of the entire ensemble to remain zero. 
Upon insertion in a $\sim 500$k atoms cell, CNA and DXA analyses as implemented in Ovito \cite{ovito} were used to ensure that no defects other than the two dislocation networks were introduced in the simulation volume.

\subsection{DDD simulations} \label{sec:DDD}

DDD simulations were performed using the ParaDiS code developed and maintained at LLNL \cite{Arsenlis07}. In the DDD model, dislocation lines are discretized into a set of segments interconnected through dislocation nodes \cite{Cai06}. The dislocation system is then evolved in time by calculating nodal velocities using a mobility function which describes the local response of the dislocation nodes to the driving force. Material parameters for the DDD model were computed using the same interatomic potential model of Ta as in the corresponding MD simulation: lattice constant $a_0 = 0.33032$ nm, shear modulus $\mu = 55$ GPa and Poisson's ratio $\nu = 0.34$. We employed linear mobility functions both for the edge and the screw dislocations. However, the drag coefficient for the screws was more than 100 times greater than that of the edges, at $5.0 \times 10^{-2}$ Pa$\cdot$s and $3.8 \times 10^{-4}$ Pa$\cdot$s respectively. The screw mobility function was of a simple pencil type \cite{cai2004mobility, Arsenlis07}. In all cases the drag coefficient for climb motion was set to a high value of $10^4$ Pa$\cdot$s so that climb was effectively suppressed.

Central to the DDD method is the kinematics of the dislocation nodes by which motion of the dislocation network is prescribed. In the following section, kinematics of network nodes as presently implemented in our DDD code is reviewed.

\section{Kinematics of conservative motion of network 3-nodes} \label{sec:oldkinematics}

\begin{figure}[t]
  \begin{center}
    \includegraphics[width=0.48\textwidth]{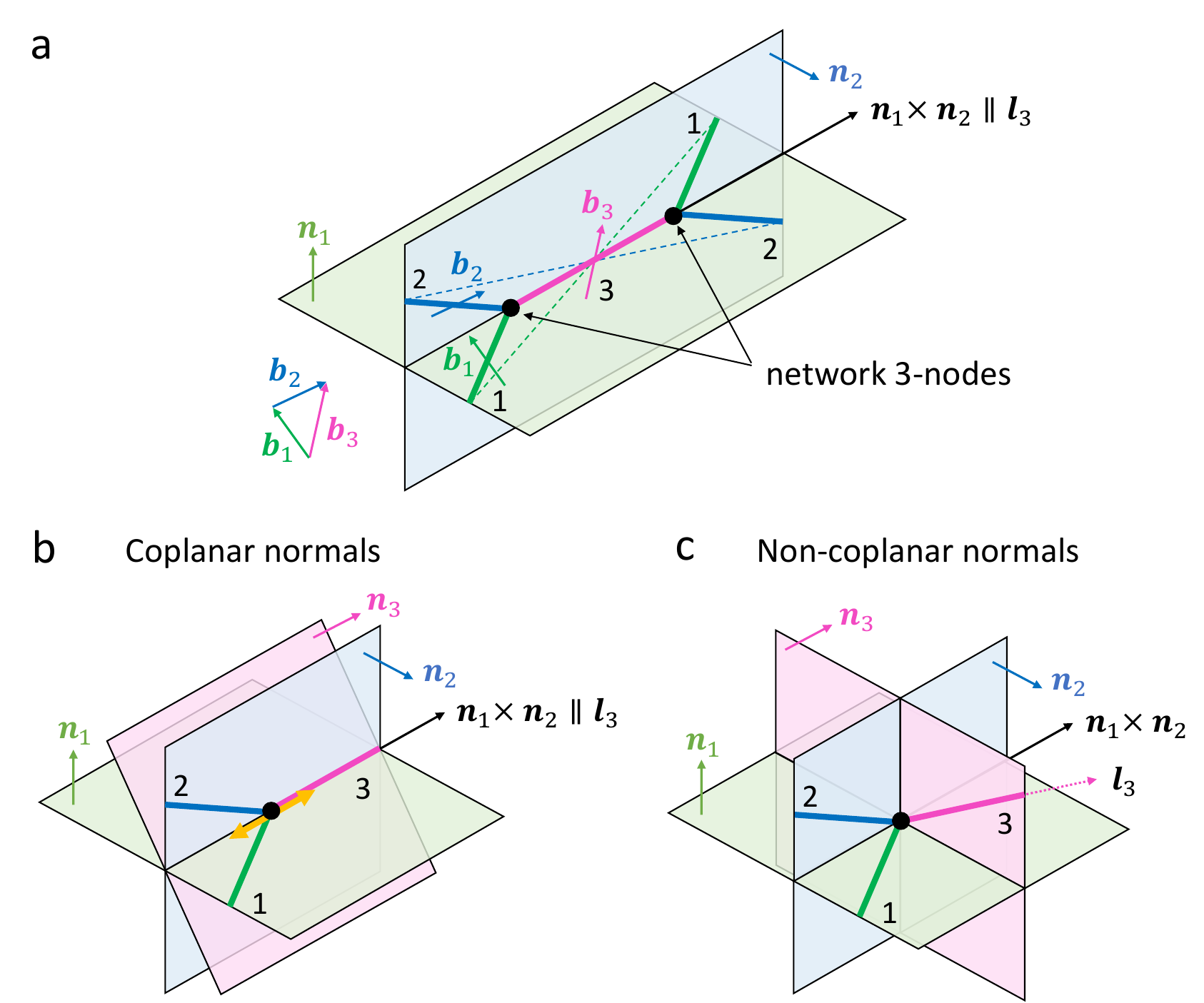}
  \end{center}
  \caption{(a) Two dislocations 1 and 2 with Burgers vectors $\vect{b}_1$ and  $\vect{b}_2$ move towards each on their geometric glide planes defined by normals $\vect{n}_1$ and $\vect{n}_2$. After initial intersection (dashed blue and green lines) a product junction dislocation 3 (thick magenta line) with Burgers vector $\vect{b}_3 = \vect{b}_1 + \vect{b}_2$ zips along line $\vect{l}_3$ (solid black line) at the intersection of planes 1 and 2, $\vect{l}_3 \parallel (\vect{n}_1 \times \vect{n}_2)$. Geometric glide plane of the product dislocation is defined by normal $\vect{n}_3 = (\vect{l}_3 \times \vect{b}_3)$ which is co-planar with normals $\vect{n}_1$ and $\vect{n}_2$. The product dislocation can glide away from the initial zipping line $\vect{l}_3$ but must remain in its geometric glide plane with normal $\vect{n}_3$. 
  (b) Focusing on one of two 3-nodes formed in dislocation reaction depicted in (a), the 3-node can glide conservatively only along the intersection line $\vect{l}_3$.
  (c) In our MD and DDD simulations we observe numerous 3-nodes in which normals of three glide planes are non-coplanar. Consequently, none of the three dislocations entering the node, including the junction line $\vect{l}_3$ (dashed magenta line), is parallel to the intersection of glide planes of two other dislocations (solid black lines). According to the basic kinematic rule previously implemented in our DDD model, such sticky 3-nodes can not move conservatively. 
  }
\label{fig:binary_junction}
\end{figure}

Our work concerns dislocation junction nodes that form when two dislocation lines -- parents -- collide and merge with each other resulting in zipping a third common line, a product or junction dislocation.
Forming at the junction's ends is a network 3-node in which two parent dislocations and the product dislocation connect together (Fig.~\ref{fig:binary_junction}).
Lying at the intersection of the parent glide (habit) planes the associated network 3-nodes have been assumed to act as strong pinning points hindering further motion of all three dislocations resulting in an increase in the flow stress, the phenomenon referred to as strain hardening.
Furthermore, junction 3-nodes have been predicted to enhance dislocation multiplication~\cite{madec2002dislocation, devincre2008dislocation, stricker2015dislocation, SillsBertinAghaeiCai} thus further adding to strain hardening.

Before describing our findings, it is useful to first review the kinematics of nodal motion as presently described in the literature (e.g. \cite{cai2004mobility, BulatovCai06}) and implemented in our ParaDiS DDD model and code \cite{Arsenlis07}. 
Based on extensive literature search and communications with DDD practitioners, it is our understanding that most of existing 
DDD models presently in use rely on similar if not identical rules.
Consider the schematic in Fig.~\ref{fig:binary_junction} where dislocation lines numbered $i = $ 1, 2 and 3 merge in a 3-node with their three Burgers vectors $\vect{b}_i$ and unit vectors $\vect{l}_i$ defining their tangent line orientations numbered accordingly. A dislocation with Burgers vector $\vect{b}$ and line tangent vector $\vect{l}$ can move conservatively, i.e. not requiring any diffusional mass transport, in its geometric glide plane defined by the plane normal $\vect{n} = (\vect{b} \times \vect{l})$. Except for conditions not considered here, conservative motion or \emph{glide} is much easier than any non-conservative motion such as \emph{climb} that takes dislocations out of their glide planes. 
Expressed algebraically, for a dislocation to move conservatively its velocity vector $\vect{v}$ should satisfy the condition 
$\vect{n} \cdot \vect{v} = 0$.
For a 3-node to be able to glide conservatively with velocity $\vect{v}$,
the same condition should be simultaneously satisfied for all three lines merging at the node
\begin{equation}
\vect{n}_i 
\cdot \vect{v} = 0 ; ~i = 1, 2, 3.
  \label{eq:kinematic_rule}
\end{equation}
Thus, a network 3-node is permitted to glide or move conservatively in the subspace complementary and orthogonal to the space spanned by the three normal vectors.
Because $\vect{n}_i = (\vect{b}_i \times \vect{l}_i)$, it is worth pointing out that Eq.~(\ref{eq:kinematic_rule}) pertaining to the glide constraint of dislocation line $i$ is automatically satisfied if the node moves either along the Burgers vector direction ($\vect{v} \parallel \vect{b}_i)$ or along the line direction ($\vect{v} \parallel \vect{l}_i)$.

The concise rule expressed by Eq.~(\ref{eq:kinematic_rule}) seemingly covers all relevant cases of conservative motion of a 3-node. 
When the three normals are linearly independent of each other (the normals are non-coplanar), the complementary motion subspace is 0-D meaning that the 3-node is not permitted to glide conservatively, Fig.~\ref{fig:binary_junction}c.
When the normals span a 2-D subspace (the normals are co-planar), the permitted complementary subspace is 1-D so that the node can move along a line parallel to all three glide planes. This covers the most frequently encountered situation depicted in Fig.~\ref{fig:binary_junction}b when one of three dislocations is a junction dislocation zipped in a collision of two parent dislocations gliding conservatively on their respective glide planes. 
When the normals are all parallel to each other, the motion subspace is 2-D and the 3-node can move conservatively in the common glide plane of three constituent dislocations, as in planar dislocation networks.  
The same rule applies when one or more dislocations are pure screw. When the Burgers vector is exactly parallel to the line direction vector as in a screw dislocation, $\vect{n} = \vect{0}$ meaning that no unique geometric glide plane can be defined -- a screw dislocation can indeed glide conservatively in any plane containing its Burgers vector. Since its normal vector is zero, the basic glide constraint condition is satisfied for any velocity vector meaning that a screw dislocation contributes no constraint on the 3-node glide. 
In particular, when all three dislocations are screws, the subspace spanned by three (zero) normals is 0-D meaning that the node can glide unconstrained in 3-D~\cite{bulatov2002nodal}.  Parenthetically, the same concise rule defines the subspace for conservative motion of dislocation network nodes of arbitrary degree ($\geq 3$).

\section{Results}

\subsection{X-scale observations} \label{sec:Xscale}

\begin{figure}[!t]
  \begin{center}
    \includegraphics[width=0.6\textwidth]{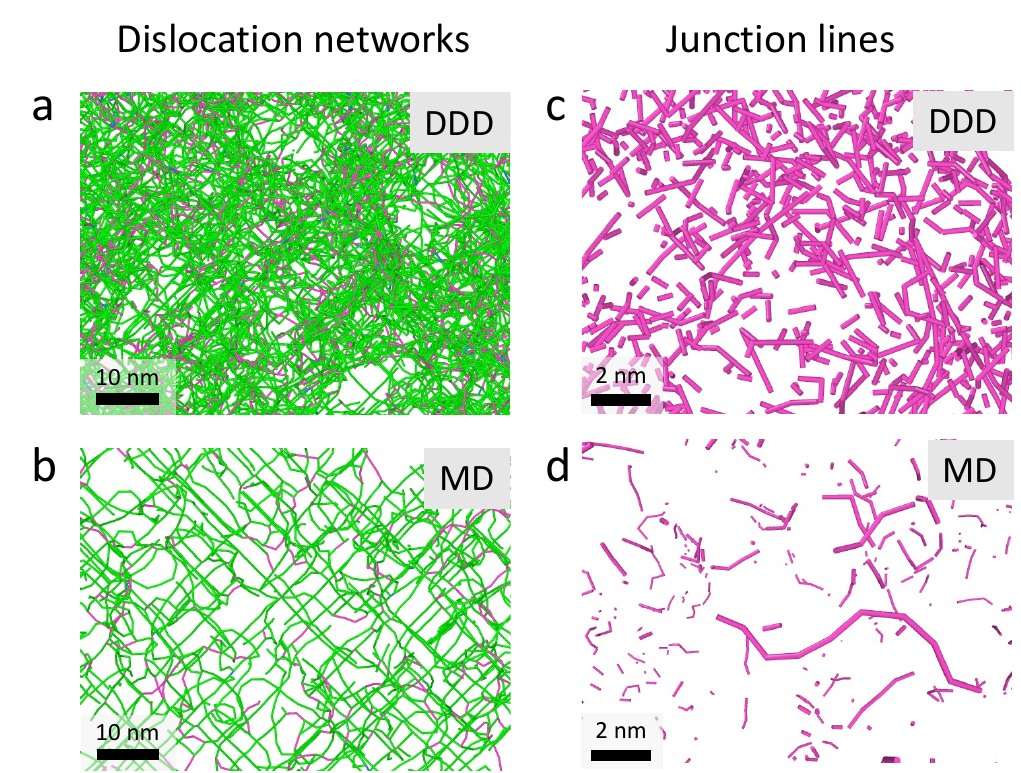}
  \end{center}
  \caption{Snapshots of dislocation networks in BCC Ta at strain 0.4. (a) In a DDD simulation dislocations develop persistent networked tangles that grow increasingly dense under continued straining. (b) In an MD simulation performed under identical conditions dislocation tangles also form but are much less dense and do not persist. (c-d) Representative snapshots in which only the $\langle 100 \rangle$ junction dislocations are shown while all $ \frac{1}{2} \langle 111 \rangle$ dislocations are deleted. (c) In DDD, dislocation junctions present themselves as short line segments that remain straight and static throughout the simulation. (d) In MD, the same $\langle 100 \rangle$ junction dislocations can grow considerably longer and often develop bends and corners.}
\label{fig:md_ddd_xscale}
\end{figure}

In comparing our MD and DDD simulations under identical conditions, we observed considerable discrepancies in the rates of dislocation multiplication and strain hardening.
In MD simulations we consistently observe flow stress, dislocation density and other characteristics of material response to attain stationary values and to stay subsequently unchanged provided straining conditions are maintained unchanged over a sufficiently long time \cite{Zepeda17, bertin2022sweep, STIMAC2022}.
DDD simulations, however, predict a gradual buildup of increasingly dense dislocation tangles that never dissolve (Fig.~\ref{fig:md_ddd_xscale}a and Movie~S1) and result in seemingly permanent storage of immobile dislocations and unending hardening (e.g. see Fig.~\ref{fig:md_ddd_density}).
In corresponding MD simulations dislocation tangles also form but are much less dense and sooner or later dissolve and do not persist throughout the simulation (Fig.~\ref{fig:md_ddd_xscale}b and Movie~S2).

So what holds the dense tangles together leading to ostensibly permanent storage of dislocations in DDD simulations?
Exploratory parametric studies suggested that a state of steady flow of the kind observed in MD can be attained in a DDD simulation, but at the cost of raising climb mobility of dislocation segments well above physically justifiable levels. Climb motion violates the conservative glide constraints and is controlled by slow diffusional mass transport. On a few nanoseconds time scales of our MD simulations, climb motion is exceedingly unlikely and can be ruled out. Further zeroing in on climb, we observed that, for steady flow to become attainable in a DDD simulation, it was sufficient to artificially raise climb mobility only for dislocation segments directly connected to 3-nodes while keeping climb mobility of all other dislocation segments near zero.

\begin{figure}[!t]
  \begin{center}
    \includegraphics[width=0.6\textwidth]{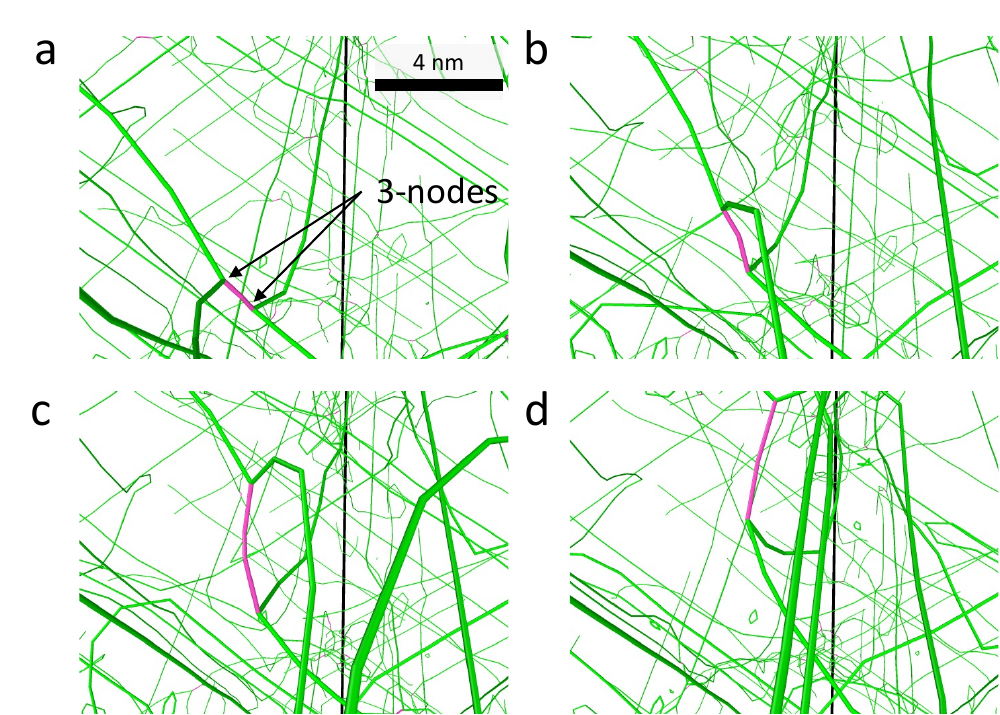}
  \end{center}
  \caption{A typical sequence of 3-node motion observed in a large-scale MD simulation in BCC Ta: going from frame (a) to frame (d) two 3-nodes at the ends of a binary junction line (magenta) experience considerable motion. More clearly observed in a detailed visualization of the same sequence in Movie~S3, the junction 3-nodes appear to be more mobile that some of the surrounding dislocations. Resulting from the motion of 3-nodes the junction dislocation changes its line orientation tracing a curved path in space. The total elapsed time from frame (a) to frame (d) is $10$ ps. }
\label{fig:3node_motion}
\end{figure}

So do the tangles dissolve in MD simulations due to hard to explain elevated climb mobility of dislocation segments entering the 3-nodes? 
An essential clue eventually leading us to solving this puzzle came in videos of dislocation network evolution extracted from MD simulations. One typical sequence is shown in Fig.~\ref{fig:3node_motion}a-d and Supplementary Movie~S3, in which some of the 3-nodes are observed to move along distinctly curved paths in space. 
As a result, many junctions appearing in MD simulations become curved, developing bends and even sharp corners (Fig.~\ref{fig:md_ddd_xscale}d). This is in contrast to abundant perfectly straight junction dislocations observed inside dense dislocation tangles in DDD simulations (Fig.~\ref{fig:md_ddd_xscale}c). 
In principle, a 3-node can move along such a curved trajectory if one or more dislocations entering the node are screws or, short of that, if some of participating dislocations move by climb thus violating the glide constraints expressed in the kinematic rule, Eq.~(\ref{eq:kinematic_rule}).  Dismissing climb motion as not viable, careful analysis of 3-node geometries immediately preceding formation of characteristic bends on the junction lines revealed that in most such instances none of the three dislocations segments were screw. Instead, immediately preceding instances when a junction bends, the 3-node is observed to attain a configuration in which the normals to the three geometric glide planes become linearly independent, i.e. span the 3-D space, just as in the hypothetical configuration in Fig.~\ref{fig:binary_junction}c. According to the kinematic rule of nodal motion implemented in our DDD model, Eq.~(\ref{eq:kinematic_rule}), such 3-nodes are restricted from moving conservatively and indeed immobile; hence we refer to them as sticky nodes.  Yet in the counterpart MD simulations, such 3-nodes are observed to start moving nearly instantly after attaining a sticky geometry and to move briskly, most often sideways relative to the junction dislocation line direction.

\begin{figure}[t]
  \begin{center}
    \includegraphics[width=0.6\textwidth]{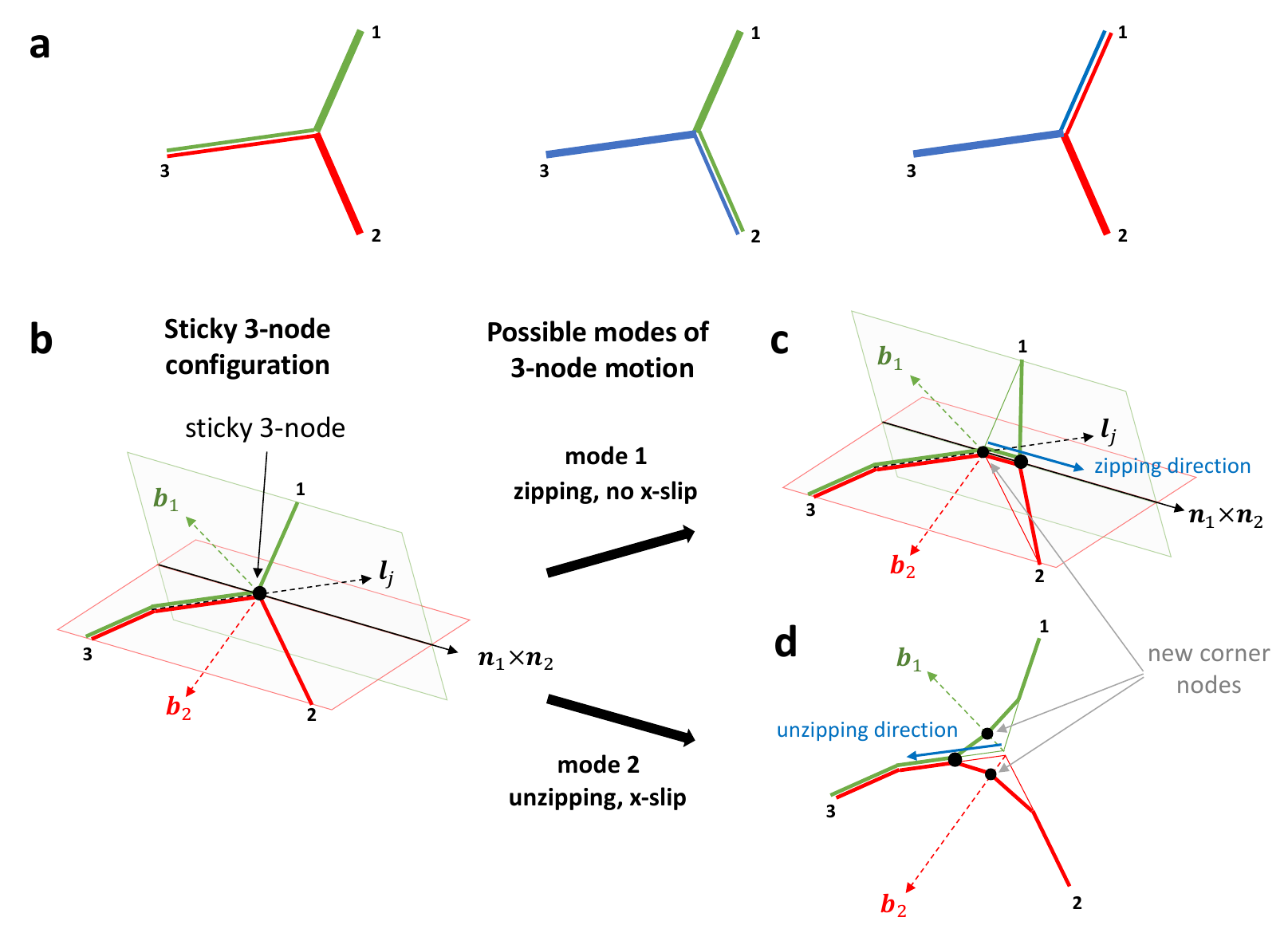}
  \end{center}
  \vspace*{-0.5cm}
  \caption{Mechanisms of coordinated conservative motion of an initially sticky 3-node. (a) Three alternative representations of a single dislocation network 3-node. (b) An initially sticky 3-node may undergo two possible modes of motion depending on the driving force at the node: (c) Mode 1 entails zipping line 3 along the intersection of glide planes 1 and 2 shown as a solid black line. Glide planes 1 and 2 are also shown semi-transparent in colors matching the active lines. In this example, the additional zipping creates a bend on junction line at the position of the original sticky 3-node. (d) Mode 2 entails unzipping line 3 by coordinated motion of dislocations 1 and 2. Two corner nodes delineating line segments in the original and the cross-slip planes are shown as black dots. In both modes, topological rearrangement turns the original sticky 3-node into a regular 3-node.
  }
\label{fig:3node_kinematics}
\end{figure}

\subsection{Mechanisms of enhanced nodal mobility}

Fig.~\ref{fig:3node_kinematics} explains how a sticky 3-node is resolved in a purely conservative motion and why such nodes often move sideways.  
As depicted in Fig.~\ref{fig:3node_kinematics}a, it is possible and useful for our purposes here to view one of three dislocations merging in a 3-node as a passive or product dislocation (junction) formed by zipping together two active (parent) dislocations. 
For each 3-node there are three equivalent ways to define a parent pair and their product and, hence, three representations of the same 3-node: $1+2\to3$, $1+3\to2$, and $2+3\to1$. 
In the following we consider the the first of three representations, i.e., $1+2\to3$, but the same considerations are applicable when dislocation 2 or dislocation 3 are viewed as the reaction products of dislocations 1 and 3 or 1 and 2, respectively.

Fig.~\ref{fig:3node_kinematics}c depicts a mechanism in which a sticky 3-node moves by way of two dislocations 1 and 2 zipping a new segment of the passive junction dislocation 3. The newly drawn junction segment is not co-linear with the previously existing junction thus forming a bend on the junction line. The mechanism is fully conservative despite such 3-node classified as non-coplanar or sticky in our DDD simulations according to the kinematic rule (Eq.~\eqref{eq:kinematic_rule}). The seeming contradiction is resolved by observing that active dislocations 1 and 2, while zipping a new segment of junction, preserve their glide constraints whereas the junction dislocation itself is passively extended (zipped) but otherwise is staying motionless thus placing no constraint on the nodal motion. The glide plane of the newly drawn junction segment $\vect{l}'_3$ is now defined by normal vector $\vect{n}'_3 = (\vect{b}_3 \times \vect{l}'_3)$, which is co-planar with glide plane normals $\vect{n}_1$ and $\vect{n}_2$ of two active dislocations and different from the non-coplanar normal $\vect{n}_3$ of the junction dislocation in the sticky configuration prior to zipping.  
The glide constraint $\vect{n}'_3 \cdot \vect{v} = \vect{0}$ is automatically satisfied because $\vect{v} \parallel \vect{l}'_3$ as discussed earlier.
The bend forming on the junction line marks the position in which the 3-node was sticky. After the zipping mechanism has taken place, the 3-node is no longer sticky.

Fig.~\ref{fig:3node_kinematics}d depicts another mechanism by which a 3-node can move conservatively starting from the same sticky configuration in Fig.~\ref{fig:3node_kinematics}b. Here, rather than zipping a new junction segment, two active lines move along and unzip the passive junction dislocation. Because the 3-node is sticky, unzipping the junction along vector $\vect{l}_3$ would seemingly violate the base kinematic rule for conservative motion. The impasse is resolved by observing that glide planes of two active lines do not need to be preserved for the 3-node to move conservatively by unzipping. The sticky node can indeed glide if two active lines somehow transfer themselves -- cross-slip -- from their initial glide planes that are not parallel to vector $\vect{l}_3$ into new glide planes that are parallel to the same vector. The new glide planes -- the cross-slip planes -- are defined by normal vectors $\vect{n}_i' = (\vect{b}_i \times \vect{l}_3); ~ i = 1,2$. As shown in the schematic in Fig.~\ref{fig:3node_kinematics}d, cross-slip can take place at the cost of creating two corner cross-slip 2-nodes, one corner node for each active line. Each corner node can glide conservatively, but only along the line parallel to the corresponding Burgers vector $\vect{b}_i$ emanating from the initial (sticky) position of the 3-node. Simultaneously, the 3-node itself can glide conservatively along the junction line. 
Furthermore, even when the 3-node is sticky, it is still possible for the cross-slip plane of one of the two active dislocations to coincide with the original glide plane of the same dislocation. In such degenerate cases, only one of the two active dislocations has to cross-slip for the 3-node to move conservatively by unzipping.
Here again, the 3-node is no longer sticky after such an unzipping event has occurred.

The unzipping mechanism illustrated in Fig.~\ref{fig:3node_kinematics}d is just one manifestation of a previously unaccounted for wider class of mechanisms of nodal mobility we here term {\it nodal cross-slip}, by which one or more dislocations entering a network node change their glide plane(s).  Unlike conventional cross-slip that entails change in a glide plane of a screw dislocation, nodal cross-slip mechanism of the kind depicted in Fig.~\ref{fig:3node_kinematics}d does not require any of the three participating dislocations to be screw. Rather unexpectedly, by virtue of their connectedness into a network at 3-nodes, dislocations acquire additional freedoms of conservative motion -- nodal cross-slip -- that neither one of them can afford individually. 
In another series of MD simulations to be reported elsewhere we observe and analyze a completely different mechanism of nodal cross-slip taking place in the face-centered cubic (FCC) metal aluminum.

The common kinematic rule Eq.~(\ref{eq:kinematic_rule}) requires glide planes of all three dislocations to be preserved simultaneously. As such it is a sufficient, but not a necessary condition for a conservative motion of a 3-node. In infrequent but important instances when a 3-node becomes sticky, the same rule becomes overly restrictive. Our two newly discovered mechanisms (Figs.~\ref{fig:3node_kinematics}c-d) show that all three glide planes do not need to be simultaneously preserved and that instead a sticky 3-node can always be resolved 
-- i.e. moving while satisfying all glide constraints --
and even possibly in several distinct directions, at a cost of one, two or even possibly all three dislocations having to change their glide planes.  Once resolved, a sticky 3-node becomes a regular 3-node (with co-planar glide plane normals) that can now move conservatively while obeying the same kinematic rule in which however some of the glide plane normals have changed.

\subsection{Verification of zipping and unzipping mechanisms} \label{sec:network}

\begin{figure*}[t]
  \begin{center}
    \includegraphics[width=0.95\textwidth]{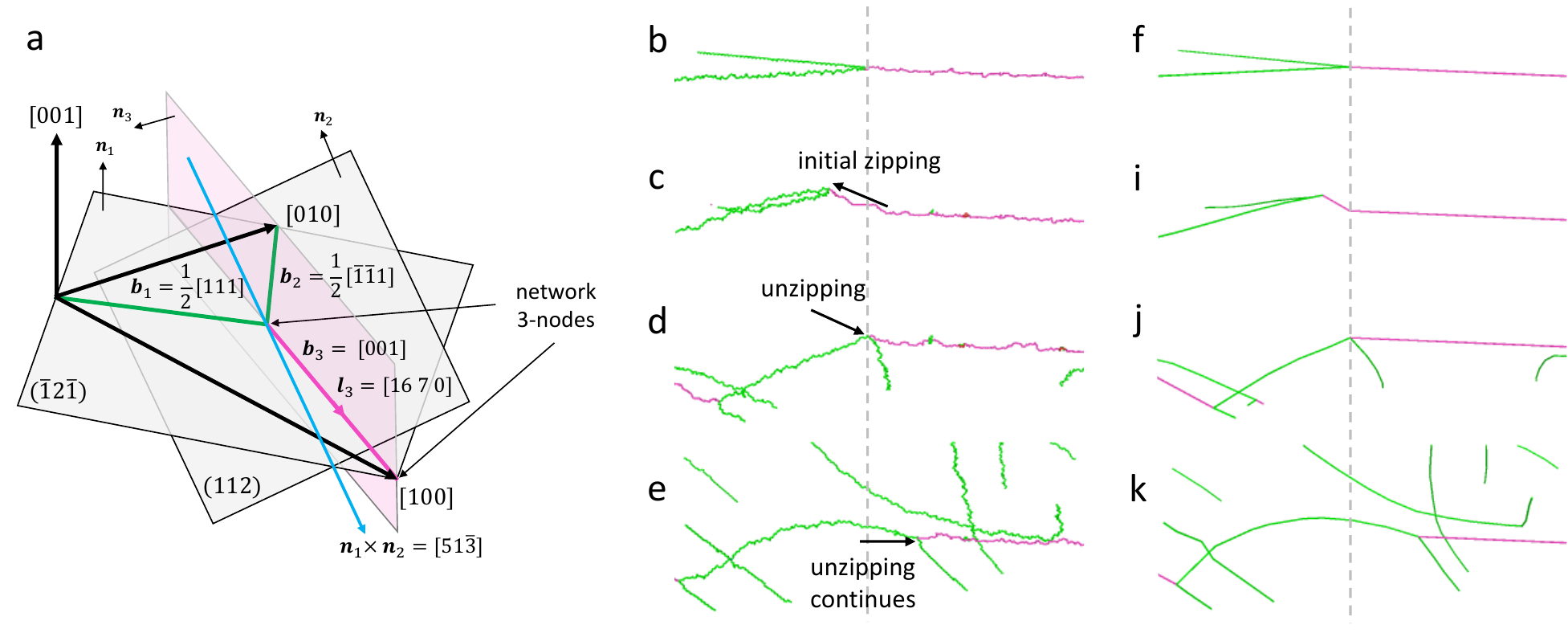}
  \end{center}
  \vspace*{-0.5cm}
  \caption{Geometry and evolution of an elemental dislocation network with a single 3-node with non-coplanar glide plane normals. (a) The network is made up of two $\frac{1}{2}\langle 111 \rangle \{112\}$ dislocations (green) of non-screw (mixed) characters and a $\langle 001 \rangle$ dislocation (magenta). As inserted, the 3-node is sticky since normal vectors $[\bar{1}2\bar{1}]$, $[112]$ and $[\bar{7}, 16, 0]$ of the glide planes of its three constituent dislocations  are non-coplanar. (b-e) Snapshots of a MD simulation in BCC Ta. (b) 3-node configuration right after it is inserted in the crystal. Two dashed gray vertical lines are drawn to show initial positions of the sticky 3-node in the MD and the DDD simulations respectively. (c) Configuration attained after initial relaxation at 300K under zero stress. The $\langle 001 \rangle$ dislocation additionally zips along the intersection of glide planes of two $\frac{1}{2}\langle 111 \rangle$ dislocations and forms a bend at the initial position of the 3-node. (d) Upon subsequent application of tensile strain, the newly zipped portion of the $\langle 001 \rangle$ dislocation gradually unzips and the 3-node returns to its sticky state at the bend. (e) On further straining, unzipping of the $\langle 001 \rangle$ line continues past the bend meeting no resistance. (f-k) Snapshots from a DDD simulation with augmented kinematics performed under conditions identical to the MD simulation. (f) Initial network configuration. (i) After relaxation. (j-k) After applying tensile straining.  Overall, network evolution in the DDD simulation shown in (f-k) matches the MD simulation shown in (b-e).}
\label{fig:md_ddd_elemental}
\end{figure*}

To confirm that the two mechanisms described above do indeed enable conservative motion of sticky 3-nodes, we performed MD simulations of a single crystal body-centered cubic (BCC) tantalum containing a pair of elemental dislocation networks, see \S\ref{sec:MD}.
For clarity, only one of the two elemental networks is shown in Fig.~\ref{fig:md_ddd_elemental}.
The network is constructed so as to contain a sticky 3-node connecting two $\frac{1}{2}\langle 111 \rangle$ dislocations and one $\langle 100 \rangle$ dislocation.
Fig.~\ref{fig:md_ddd_elemental}a shows the geometry in which all three glide plane normals $\vect{n}_i$ of the segments connected to the 3-node are linearly independent. In this geometry none of three line tangent directions is parallel to the intersection of glide planes of the other two dislocations.  

At the first stage, after inserting it into the crystal the elemental dislocation network was annealed at $300$~K under zero applied stress. During annealing the $\langle 100 \rangle$ junction dislocation spontaneously extends by zipping two $\frac{1}{2}\langle 111 \rangle \{112\}$ parent dislocations along the $[5 1 \bar{3}]$ direction common to their two glide planes which is different from the initial as-inserted line direction $[16\,7\,0]$ of the same junction, Fig.~\ref{fig:md_ddd_elemental}c. Junction extension past the sticky 3-node proceeds precisely by the purely conservative zipping mechanism shown in Fig.~\ref{fig:3node_kinematics}c in which the junction acquires a corner of the kind we frequently observe in our large-scale MD simulations in the same model of tantalum. During and to the end of this zipping motion, the 3-node is no longer sticky as its three glide-plane normal vectors of three segments connected to the 3-node span a 2-D space.

At the second stage, after annealing, the crystal was strained along the [001] direction at a strain rate of $2\times 10^{8}\,{\rm s}^{-1}$.  During straining, the junction segment  gradually unzips (Fig.~\ref{fig:md_ddd_elemental}d-e) causing the 3-node to move back to its initial position before annealing. Unzipping proceeds in the exact reverse order to the preceding zipping motion by the conventional conservative mechanism during which the 3-node remains regular (not sticky).  
On reaching its original position before annealing the 3-node becomes momentarily sticky again, but continues its motion with no visible hesitation by unzipping the junction dislocation along its as-inserted direction $[16\,7\,0]$. Unzipping past the sticky configuration proceeds by  conservative unzipping mechanism of nodal cross-slip schematically depicted in Fig.~\ref{fig:3node_kinematics}d in which two $\frac{1}{2}\langle 111 \rangle \{112\}$ transition into their cross-slip planes.

\subsection{Implementation of new mechanisms of mobility in DDD} \label{sec:DDDimplementation}

Both zipping and unzipping modes of sticky node motion shown in Fig.~\ref{fig:md_ddd_elemental}c and \ref{fig:md_ddd_elemental}e are prohibited in our DDD implementation of the conventional rule for conservative motion of 3-nodes (Eq.~\eqref{eq:kinematic_rule}). 
To address this deficiency, we augmented kinematic rules of nodal motion in DDD to permit conservative motion of 3-nodes as observed in our MD simulations. For that we added special topological operators to handle dissociation of sticky 3-nodes into non-sticky 3-nodes (see \ref{sec:split3node}). Given that each 3-node permits three alternative representations (depending on which dislocation is considered inactive), our algorithm generates trial split configurations for three modes of unzipping and three modes of zipping and computes the rate of power dissipation for each of the resulting six modes (see \cite{Arsenlis07} for details). The trial mode with the highest dissipation rate is accepted and the selected splitting mode of the 3-node is executed.

\subsection{DDD simulations with amended kinematics of 3-node motion} \label{sec:results}

We first validate our augmented 3-node mobility kinematics by performing DDD simulations on the same elemental dislocation network previously simulated using MD, Fig.~\ref{fig:md_ddd_elemental}. To be consistent, the loading conditions in the DDD simulations followed precisely the same sequence as in MD -- first relaxation then uniaxial straining. The DDD model was carefully parameterized with material parameters computed for the same interatomic potential model of Ta as in the corresponding MD simulation. Two DDD simulations were performed with the same parameters, one using the old set of kinematic rules and the other using the augmented kinematics permitting conservative motion of sticky 3-nodes. In both DDD simulations, non-conservative motion by climb was effectively suppressed by setting the climb mobility coefficient to a near zero value of $10^{-4}$ (Pa$\cdot$s)$^{-1}$, such that 3-node motion can only proceed via the allowed set of kinematic rules associated with each case.

In the DDD simulation with the original kinematic rules, no motion of the sticky 3-node is observed both during relaxation and during subsequent uniaxial straining. The 3-node holds as an indestructible pinning point promoting incessant Frank-Reed dislocation multiplication. In the DDD simulation with the augmented kinematics, the evolution is qualitatively identical to that in MD.  As shown in Fig.~\ref{fig:md_ddd_elemental}i, during the initial relaxation the junction spontaneously zips past the initial 3-node position thus forming a corner on the junction line. Upon application of tensile straining, the newly formed junction segment is seen to unzip all the way to and past the corner node, after which the 3-node stops at a position close to that attained in MD, Fig.~\ref{fig:md_ddd_elemental}j. Only then the two active lines begin to bow out and to produce new dislocations, Fig.~\ref{fig:md_ddd_elemental}k. Comparing Figs.~\ref{fig:md_ddd_elemental}b-e and \ref{fig:md_ddd_elemental}f-k, a close agreement is observed overall between the MD simulation and the DDD simulation with the augmented kinematics of 3-node motion. Even if the agreement is not fully quantitative, it is qualitatively robust and holds irrespective of such details as the mobility functions. 

\begin{figure}[t]
  \begin{center}
    \includegraphics[scale=0.37]{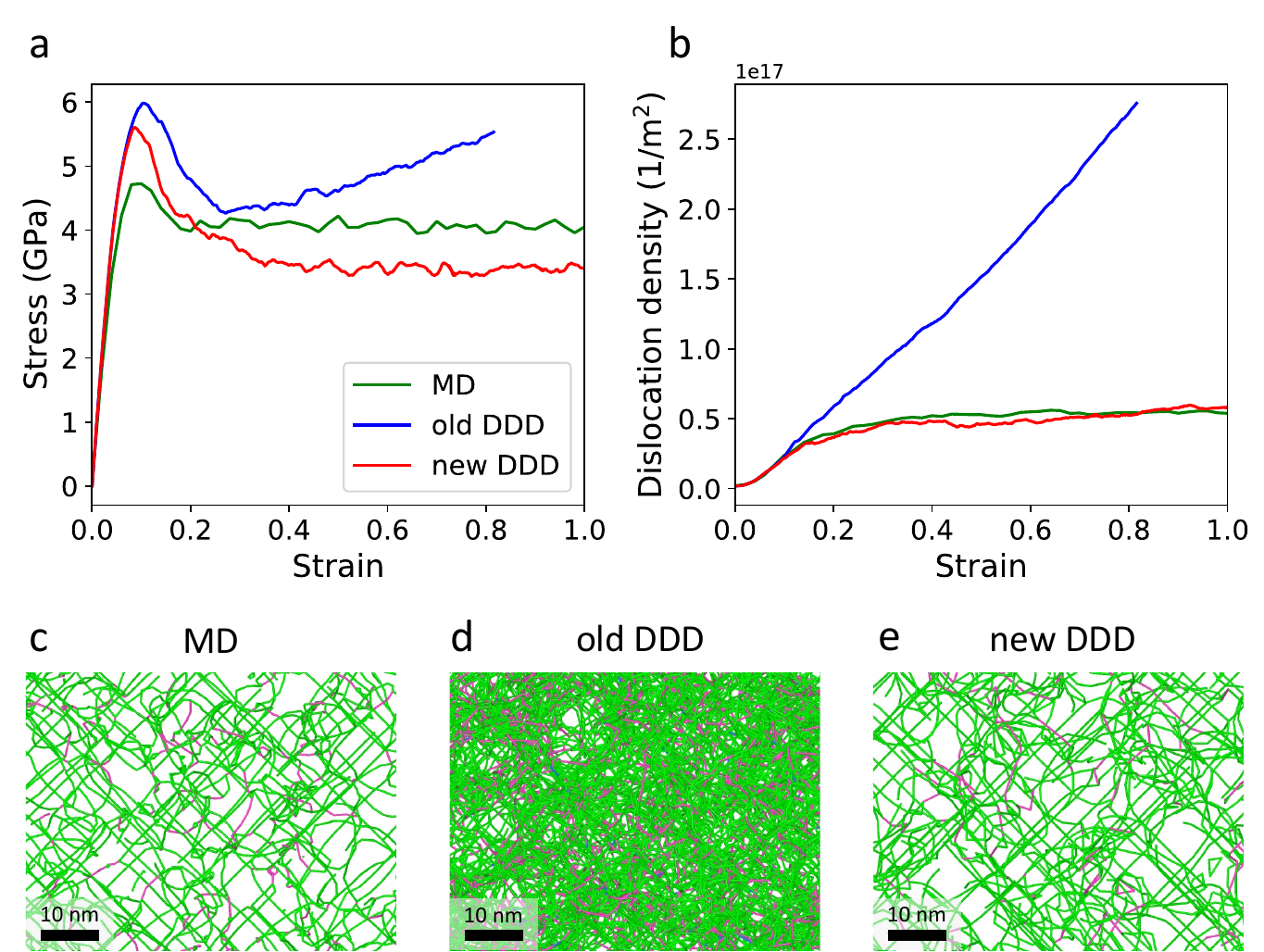}
  \end{center}
  \caption{Evolution of stress (a) and dislocation density (b) predicted in three simulations performed under identical conditions: MD simulation (green), DDD simulation with the old kinematic rules for 3-node motion (blue) and DDD simulation with the new kinematic rules permitting additional modes of 3-node motion (red). In all three simulations uniaxial compression along the [001] axis of a BCC single crystal of tantalum was performed under a constant straining rate of $2\times 10^8$ s$^{-1}$. (c-e) Snapshots of the dislocation network at strain 0.8 extracted from (c) MD, (d) old DDD and (e) new DDD.}
\label{fig:md_ddd_density}
\end{figure}

To observe the effects of augmented kinematics on macroscopic plasticity response and, in particular, on dislocation multiplication, we performed bulk DDD simulations in a material volume equal to that of a large-scale MD simulation. Just like in MD, the initial configuration consisted of 12  prismatic dislocation loops randomly seeded in the simulation volume. The crystals were then subjected to compression at the same rate of $2 \times 10^8$ s$^{-1}$ along the same [001] crystallographic axis. Evolution of dislocation density predicted in MD and DDD simulations with and without augmented rules for 3-node mobility is shown in Fig.~\ref{fig:md_ddd_density}. The two DDD simulations predict very different behaviors: whereas the simulation with the old kinematics (old DDD) predicts unending dislocation multiplication, the simulation with the augmented kinematics (new DDD) predicts saturation of dislocation density in close agreement with the MD prediction as shown in Fig.~\ref{fig:md_ddd_density}b. The flow stress attained in the new DDD simulation is lower than the flow stress observed in MD which is attributed to an over-simplified dislocation mobility function, i.e. linear pencil mobility, used in both DDD simulations. However the fact that flow stress clearly saturates in the new DDD simulation is significant and does not depend on the specific mobility function. At the same time, flow stress in the old DDD simulation is markedly higher than in MD and new DDD and, consistent with the steadily rising dislocation density, shows a tendency to continued hardening. These clear differences are attributed to spurious growth of dislocation tangles in the old DDD simulation, Fig.~\ref{fig:md_ddd_density}d, in contrast to MD and new DDD simulations in which enhanced nodal mobility dissolves dislocation tangles before they grow excessively dense, as shown in Figs.~\ref{fig:md_ddd_density}c and \ref{fig:md_ddd_density}e.

In additional DDD simulations we observe that enabling only the unzipping mechanism (mode 2) of sticky node motion seemed to be sufficient for the new DDD and MD simulations to come to a qualitative agreement. Alternatively, enabling only the zipping (mode 1) mechanism still resulted in spurious growth of dense dislocation tangles and unending strain hardening nearly identical to the old DDD kinematics. These results seemingly suggest that enabling or not of the zipping mechanism has only minor if any effect on DDD results. However, even if not explicitly implemented (or if turned off), zipping past a sticky 3-node can still be effectively achieved in DDD. This is because arms connected to a sticky 3-node can rotate around it and merge along the line of their glide plane intersection, thus effectively zipping the junction. Thus, it is not possible to fully prevent junction zipping past sticky nodes in DDD in order to gauge if and how much this mechanism affects DDD predictions. We note that zipping past sticky 3-nodes is frequently occurring in MD simulations resulting in formation of numerous bends on the junction lines. In the steady flow regime reached in our MD simulations, the dislocation microstructure, while continuously evolving dynamically, becomes statistically stationary. In such a stationary flow state dislocation junctions must form and dissolve at the same rates.  Likewise, events of junction zipping and junction unzipping past the ubiquitous sticky nodes (at junction bends) must also occur equally frequently, with the unzipping events requiring more substantial local stress to proceed.

\section{Discussion and summary} \label{sec:discussion}

Significant effects of dislocation network nodes on crystal plasticity and dislocation microstructure have been recognized for a long time, however simulation results and analyses presented in the preceding sections add an unexpected new twist to the existing understanding. Not only the network 3-nodes constrain dislocation motion by tying dislocations together, but the very same nodes furnish additional degrees of freedom for dislocation motion that individual dislocations do not possess. 
Previously unaccounted for mechanisms of nodal motion observed and rationalized in this work suggest that, despite significant body of work focused on interactions among dislocations, consequences of dislocation connectivity into a network are not sufficiently understood. A pertinent example from this work is that conservative unzipping of a sticky 3-node forces one or both active dislocations to cross-slip. This phenomenon of coordinated nodal cross-slip is principally different from a previously proposed enhancement of cross-slip rates at dislocation intersections \cite{rao2011calculations} or an enhanced rate of cross-slip on a screw dislocation connected to a regular co-planar 3-node \cite{rao2009atomistic}. To our knowledge, all earlier proposals for enhanced cross-slip focused on screw dislocations, whereas nodal cross-slip can and does take place when none of the participating dislocations is of screw or near-screw character.

\begin{figure}[t]
  \begin{center}
    \includegraphics[scale=0.3]{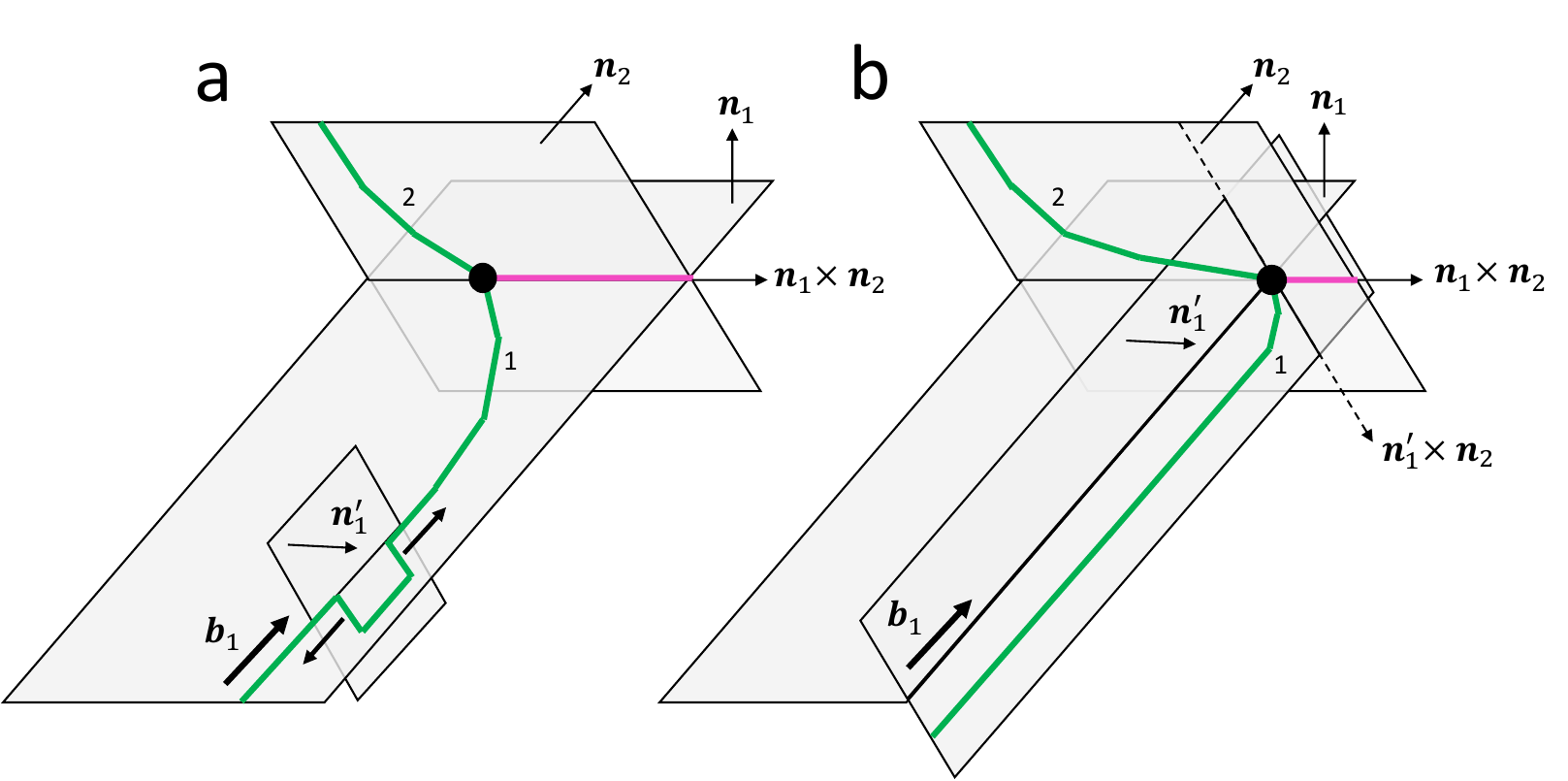}
  \end{center}
  \vspace*{-0.5cm}
  \caption{Schematic of a sequence of events commonly observed in MD simulations leading to formation of a sticky 3-node. (a) A regular (non-sticky) 3-node is formed by active lines 1 and 2 (green) gliding in planes with normals $\vect{n}_1$ and $\vect{n}_2$ and zipping a junction dislocation (magenta). One of two active lines attains a screw orientation away from the node and cross-slips from its original glide plane into a different glide plane $\vect{n}'_1$. (b) The newly formed non-screw segments of the cross-slipped dislocation move sideways with one of them impinging on the 3-node resulting in a configuration in which three dislocation segments meeting at the 3-node have linearly independent glide normals.}
\label{fig:noncoplanar_formation}
\end{figure}

Sticky network 3-nodes with three linearly independent glide plane normals are found to be an essential and pervasive component of dislocation networks observed in MD simulations of BCC tantalum. But how do such sticky 3-nodes form in the first place? 
To answer this question, we followed evolution of a few fragments of dislocation networks in very high time frame resolution. In a few instances where 3-node motion was observed to deviate from the initial straight junction line, such events were preceded by cross-slip of one of two parent $\frac{1}{2} \langle 111 \rangle$ dislocations.   As illustrated in the schematic in Fig.~\ref{fig:noncoplanar_formation}, the segment pulled into a cross-slip plane by a parent dislocation subsequently glides all the way to and impinges on the 3-node.  Thus, a regular 3-node (where the three normals span a 2-D space) initially formed by zipping two parent dislocations together, becomes sticky by virtue of one of two parent dislocations subsequently cross-slipping. Although we do not rule out that cross-slip may initiate at the 3-node itself -- as was posited in \cite{rao2011calculations} -- in a few MD sequences that we analyzed cross-slip was initiated away from the 3-nodes. It is possible that this preference reflects the known predominance of screw dislocations in BCC crystals plastically deformed at low and intermediate temperatures. Indeed, as a well-recognized consequence of a much lower mobility of screws relative to the edge dislocations, in our MD and DDD simulations alike we observe that, away from the network nodes, most dislocation lines are of screw or near-screw character.  Owing to a compact (undissociated) structure of the screw dislocation core, cross-slip of screw dislocations in BCC crystals is not a rare event of the kind observed in FCC crystals. It remains to be seen how frequently sticky nodes may form and what could be other detailed mechanisms of their formation in BCC and in materials with other crystal structures.

Behaviors and mechanisms discussed in this work were observed in atomistic simulations of BCC Ta crystals performed at very high deformation rates as was needed to reach sufficiently large plastic strains and to see a complete evolution of dislocation microstructure within the short time-scales presently accessible to MD simulations. In a first attempt to see if and how much the same mechanisms matter in crystal plasticity of other metals and at lower deformation rates, we have run additional MD simulations of the elemental dislocation network presented in Fig.~\ref{fig:md_ddd_elemental}. First, we repeated the same simulation but using another EAM interatomic potential developed for BCC W \cite{Zhou04} and observed the exact same behavior as reported in Fig.~\ref{fig:md_ddd_elemental}, i.e. zipping of the junction during relaxation followed by unzipping under loading. We also note that in another recent work \cite{bertin2023crystal} we observed pervasive bent junction lines in MD simulations of BCC Ta modeled with a SNAP potential \cite{thompson2015spectral} down to strain rates of $10^6$/s, again indicative of the prevalence of nodal cross-slip events. Thus, we expect these mechanisms of 3-node motion to be generic (at least to BCC crystals) and rather independent of the details of the interatomic potential. Second, we point out that, similar to conventional junction mechanisms, occurrence of nodal cross-slip events is governed by stress rather than by strain rate. Following this observation, we have run an additional set of simulations of the elemental network introduced in a larger box of $\sim 3.5$M atoms which we subjected to a constant stress loading. The applied stress state was determined so as to apply the same amount of shear stress on both glissile $\frac{1}{2}\langle 111 \rangle$ lines of the network while maintaining shear stress resolved on the junction line near zero. We observed the junction to unzip under a shear stress of around 400 MPa. While we believe that this value would be even lower for still longer glissile dislocation lines (i.e. using larger simulation boxes), shear stress of this magnitude is already on par with stress typically reached in BCC metals under low strain rate conditions. Thus, these simulations suggest that nodal cross-slip mechanisms should be operational under typical experimental deformation rates. Nevertheless, additional, dedicated work will be required to fully investigate the effect of these mechanisms under different conditions, including quasi-static straining, and in other crystal structures.

Even if discovered through X-scale matching of MD and DDD simulations performed on the same short time-scales, the very existence of sticky network nodes and previously unaccounted conservative mechanisms of their motion are now firmly established. Geometrical in nature, the same mechanisms must be operational on any time scales and need to be enabled for DDD simulations to be predictive. The significant improvement in the agreement between high-rate MD and DDD simulations achieved here by enabling conservative motion of sticky nodes is indicative of potentially broader importance of such mechanisms in crystal plasticity.  

In summary, we observe in MD and DDD simulations the formation of sticky dislocation 3-nodes with non-coplanar glide plane normals and report previously unknown mechanisms of enhanced mobility of such sticky nodes in BCC crystals under high strain rates. In analyzing dynamic evolution of dislocation networks in large-scale MD simulations, we observe that sticky 3-nodes are quite mobile, contrary to what has been commonly assumed and previously observed in DDD simulations. In their motion, 3-nodes are observed to follow curved trajectories forming bends and corners on the dislocation lines. Closer examination of the dynamics of elemental dislocation networks reveals two new mechanisms of fully conservative motion of such 3-nodes in which one or two participating dislocations are transferred into different glide planes. Enabling the newly observed conservative mechanisms of nodal motion in DDD code ParaDiS brings DDD simulations in close qualitative agreement with MD simulations performed under identical straining conditions. Our findings of these previously unaccounted mechanisms of dislocation motion and demonstration of their essential role in crystal plasticity showcase the value of X-scale matching in which new insights are gained when fully resolved MD simulations are compared against mesoscopic simulations performed under identical conditions on the same scales.

\section*{Acknowledgements}

This work was performed under the auspices of the U.S. Department of Energy by Lawrence Livermore National Laboratory under contract DE-AC52-07NA27344.
This work was partially supported by the U.S. Department of Energy, Office of Basic Energy Sciences, Division of Materials Sciences and Engineering under Award No. DE-SC0010412 (WC).

\section*{Author contributions}
Conceptualization: NB, WC, VB; Methodology: NB, WC, SA, AA, VB; Software: NB; Writing-original draft: NB, VB; Writing-review and editing: NB, WC, SA, VB

\section*{Competing interests}
Authors declare that they have no competing interests.


\appendix

\section{Implementation of augmented mobility in DDD} \label{sec:split3node}

\begin{figure}[t]
  \begin{center}
    \includegraphics[scale=0.45]{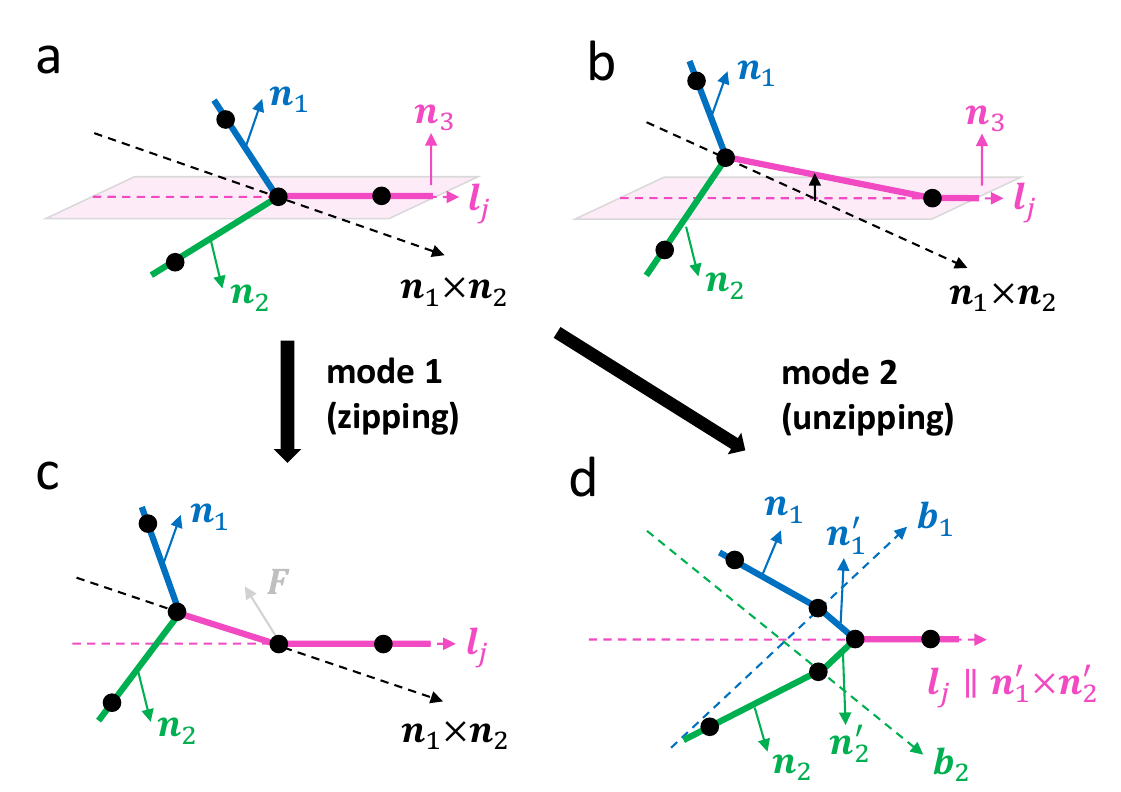}
  \end{center}
  \caption{Topological operations introduced to enable additional modes of 3-node motion. (a) Initial configuration of a 3-node with two active dislocations (green) in glide planes with normals $\vect{n}_1$ and $\vect{n}_2$ and a passive dislocation with a line direction $\vect{l}_j$ (magenta) not parallel to the intersection of two glide planes (dashed line defined by $\vect{n}_1 \times \vect{n}_2$). (b) Without adding an additional node, zipping motion of the 3-node along the dashed line changes orientation of the junction segment tilting it away from its geometric glide plane (as indicated by the black arrow), which can only take place by climb. (c) To enable mode 1 (zipping, no x-slip), the 3-node is split in two with the original node becoming a corner node and the new node moving along the dashed line. (d) In mode 2 (zipping, x-slip), the 3-node is split in three nodes with the original 3-node moving along $\vect{l}_j$ and two new corner nodes gliding along their respective Burgers vectors $\vect{b}_1$ and $\vect{b}_2$. Two short segments connecting the 3-node to the new corner nodes lie in glide planes with normals $\vect{n}'_1 \parallel ( \vect{b_1} \times \vect{l}_j )$ and $\vect{n}'_2 \parallel ( \vect{b_2} \times \vect{l}_j$ ).}
\label{fig:split3node}
\end{figure}

Kinematics of nodal motion in DDD is augmented to permit conservative motion of sticky 3-nodes as observed in our MD simulations. Although our implementation is specific to ParaDiS, the description below should be useful in suggesting appropriate modifications to motion kinematics implemented in other DDD models and codes. 

Conservative zipping of a junction connected to a 3-node is always possible but, when the 3-node happens to be non-coplanar (sticky), such zipping motion changes direction of and leaves behind a corner on the junction line.
As depicted in  Fig.~\ref{fig:split3node}b, unless the junction is indeed subdivided into two segments joined at a corner node, change in the junction zipping direction would necessitate a substantial non-conservative displacement of the junction segment out of its geometric glide plane.
In the old DDD model such a motion is not permitted, unless climb mobility is artificially elevated. Yet, this mechanism is fully conservative and does not require climb.
To address this, we added a new kinematic rule which allows a sticky 3-node to split along the junction arm via insertion of a new, initially infinitesimal junction segment. Free from the junction arm constraint, the new 3-node can then freely move along the intersection line of the glissile segments glide planes. This operation results in a corner node along the junction, and a new position for the 3-node which is no longer sticky, see Fig.~\ref{fig:split3node}c.

Similar to zipping, unzipping of a junction past a sticky node would necessitate climb unless two active dislocations are allowed to cross-slip to glide planes defined by the cross-products of their Burgers vectors with the tangent vector of the junction. As depicted in Figs.~\ref{fig:3node_kinematics}d and \ref{fig:split3node}d, cross-slip of two active lines entails formation of two cross-slip corner nodes -- one node per active line -- delineating new segments pulled into the cross-slip planes from segments moving in the original glide planes of two active dislocations. Junction unzipping beyond a sticky 3-node is purely conservative for as long as the cross-slip nodes move along the lines originating at the location of the sticky 3-node before unzipping and parallel to the Burgers vectors. In its new position the 3-node is no longer sticky.

To implement additional rules illustrated in Figs. \ref{fig:split3node}c and \ref{fig:split3node}d in ParaDiS we added two additional topological operators derived from  the original SplitMultiNode operator implemented in ParaDiS to handle dissociation of network nodes of degree four or higher, Ref. \cite{Arsenlis07}. Split3Node generates trial split configurations for three modes of unzipping and three modes of zipping. In principle, each one of the three zipping modes can proceed in either direction along an appropriate intersection line, but usually only one of two directions is aligned with the Peach-Koehler force computed on the 3-node before splitting. After six out of nine modes are selected, for each one of them an appropriate infinitesimal (in practice finite, but small) trial split node configuration is generated and its rate of power dissipation is computed (see \cite{Arsenlis07} for details). The trial mode with the highest dissipation rate is accepted and only then the so-selected change in the 3-node is executed.  

We note that, in principle, appropriate  topological operators can be added to handle dissociation of sticky network nodes of degree four and higher. Our approach is to use the SplitMultiNode operator existing in ParaDiS to test for favorable dissociations of multi-nodes into 3-nodes based on the maximum power dissipation criterion. If and when dissociation of a multi-node is deemed favored, every 3-node resulting from the accepted multi-node dissociation is then tested for a favorable zipping or unzipping motion.

\bibliography{ref}

\end{document}